\documentclass[aps,pre,twocolumn,floatfix,superscriptaddress,showpacs,showkeys]{revtex4-1}
\usepackage{epsf,amsmath,amssymb,verbatim,color,multirow,pifont,mathrsfs,soul,amsthm, wasysym}
\usepackage{graphicx, upgreek}
\usepackage[us,12hr]{datetime}

\begin{document}

\title{Scaling behaviors of information entropy in explosive percolation transitions}
\author{Yejun Kang}
\affiliation{Department of Physics, Jeonbuk National University, Jeonju 54896, Korea}

\author{Young Sul Cho}\email{yscho@jbnu.ac.kr}
\affiliation{Department of Physics, Jeonbuk National University, Jeonju 54896, Korea}
\affiliation{Research Institute of Physics and Chemistry, Jeonbuk National University, Jeonju 54896, Korea}

\begin{abstract}
An explosive percolation transition is the abrupt emergence of a giant cluster at a threshold 
caused by a suppression of the growth of large clusters. In this paper, we consider the information entropy of the cluster size distribution, which is the probability distribution for the size of a randomly chosen cluster.
It has been reported that information entropy does not reach its maximum at the threshold in explosive percolation models, a result seemingly contrary to other previous results that the cluster size distribution shows power-law behavior and the cluster size diversity (number of distinct cluster sizes) is maximum at the threshold.
Here, we show that this phenomenon is due to that the scaling form of the cluster size distribution is given
differently below and above the threshold.
We also establish the scaling behaviors of the first and second derivatives of the information entropy near the threshold to explain why the first derivative has a negative minimum at the threshold and the second derivative diverges negatively (positively) at the left (right) limit of the threshold, as predicted through previous simulation.
\end{abstract}

\maketitle

\section{Introduction}
Percolation is a phenomenon in which a giant (percolating) cluster emerges
as the link occupation probability between individuals exceeds a finite threshold~\cite{stauffer}.
The sol--gel transition~\cite{sol-gel} and metal--insulator transition~\cite{metal-insulator}
provide examples of percolation phenomena in physical systems.
Apart from the giant cluster, the distribution of finite-sized clusters also changes as the link occupation probability increases.
The uncertainty of the distribution of finite-sized clusters has been studied using percolation models
in various aspects~\cite{hassan1, hassan2, vieira, entropy_latticeanimal1, entropy_latticeanimal2, jdnoh, hklee}.
In some of these results, it has been reported that the cluster size diversity, or information entropy~\cite{shannon},
of the cluster size distribution reaches its maximum value at the threshold 
in (ordinary) random percolation models~\cite{vieira, entropy_latticeanimal1, entropy_latticeanimal2, jdnoh, hklee}.

Specifically, in~\cite{vieira}, the information entropy of the cluster size distribution 
in the Erd\"{o}s--R\'enyi (ER) model~\cite{er} and in explosive percolation (EP) models~\cite{science, dacosta_prl, dacosta_pre, dacosta_pre_numeric} 
was studied.

In the ER model, $N$ nodes are isolated at the beginning with $t=0$.
At each time step $t \rightarrow t + 1/N$, a pair of randomly selected nodes are connected by a link. 
Then, a giant cluster emerges at $t=t_c=0.5$ continuously.
On the other hand, in each EP model, 
a rule that suppresses the growth of large clusters is applied
when a pair of nodes to be connected by a link is selected for each time step~\cite{science, kahng_review, jan_review, suppression, local, choi, chae, largest, bfw}.
As a result, a giant cluster emerges abruptly (explosively) at a delayed threshold, $t = t_c>0.5$.
Numerous studies have clarified the transition nature of the EP models~\cite{fortunato, friedman, fss_exp, ziff}, and as a result,
it has been concluded that an abrupt but continuous transition occurs when a local suppression rule is applied~\cite{dacosta_prl, dacosta_pre, dacosta_pre_numeric, jan, jan2, grassberger, riordan, cho_science}, 
whereas a discontinuous transition occurs when a global suppression rule is applied~\cite{ep_growing, cho_science, riordan, jan, hybrid}.

In these models, the cluster size distribution $p_s(t) = A(t)n_s(t)$ for $n_s(t) = N_s(t)/N$ and $A(t) = 1/(\sum_{s=1}^{\infty}n_s(t))$
has been considered, where $N_s$ is the number of clusters of size $s$. 
Therefore, $p_s$ is the probability that the size of the randomly selected cluster is $s$.
We note that $\sum_{s=1}^{\infty}p_s=1$. 
Then, the information entropy of the cluster size distribution $p_s$ is given by
\begin{equation}
H(t) = -\sum_{s=1}^{\infty}p_s(t){\text {log}}_2p_s(t).
\label{Eq:Hdef}
\end{equation}

\begin{figure}[t!]
\includegraphics[width=0.85\linewidth]{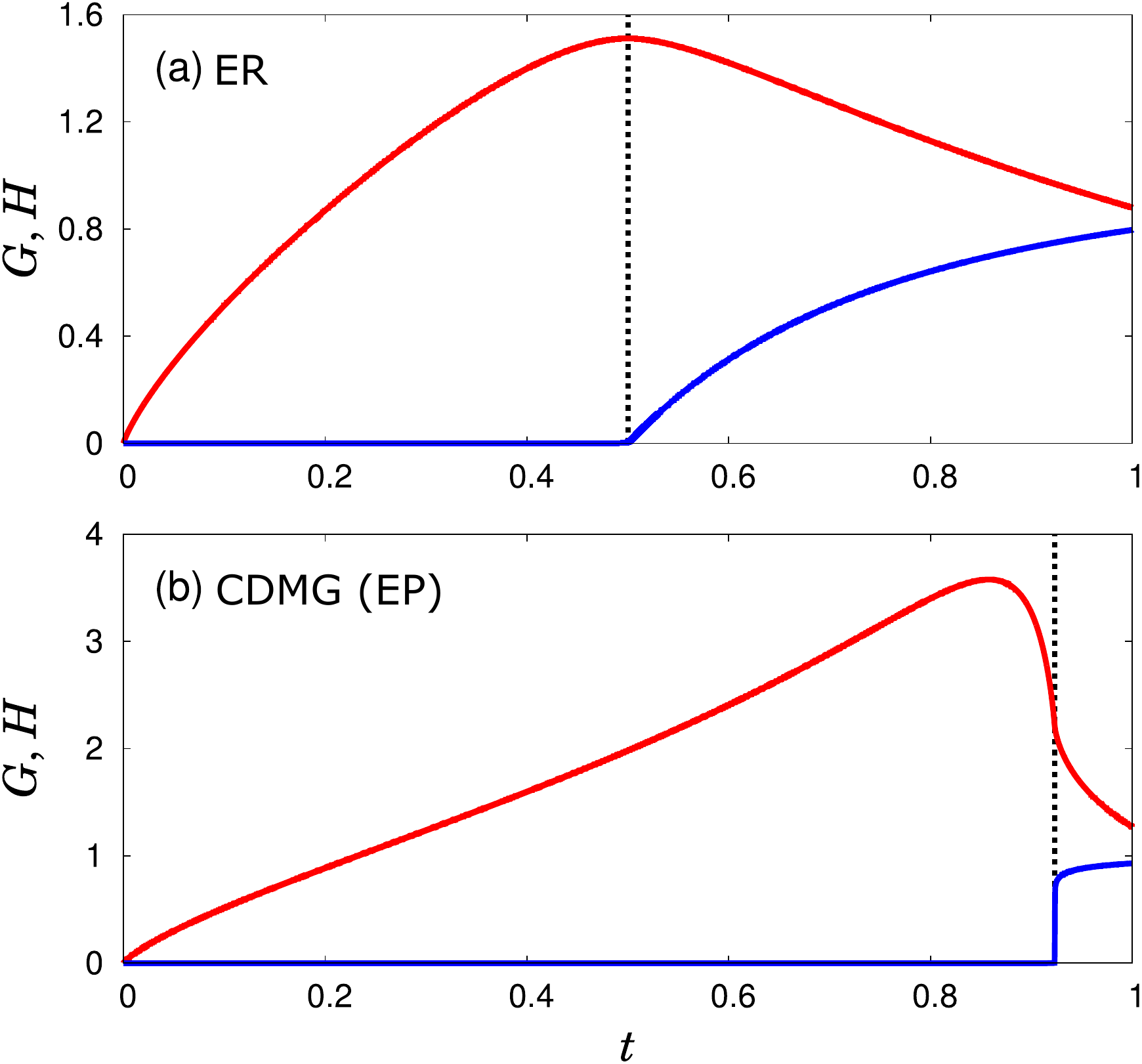}
\caption{$G$ (lower line) and $H$ (upper line) vs. $t$ in (a) the ER model and 
(b) the CDMG model with $m=2$ (see Sec.~\ref{sec:intro_dCR} for model definition) obtained by simulation with $N=1.024 \times 10^7$,
where $G$ is the fraction of nodes belonging to the largest cluster. 
In both plots, the location of the dotted line is $t_c$.} 
\label{Fig:ER_EP_H_G}
\end{figure}

Interestingly, in~\cite{vieira}, it was reported that the maximum point of $H(t)$ is equal to $t_c$ in the ER model,
whereas that of $H(t)$ is less than $t_c$ in the EP models.
These contrasting results are presented in Fig.~\ref{Fig:ER_EP_H_G}, where the rule developed by da Costa {\it {et al.}} (CDMG)~\cite{dacosta_prl, dacosta_pre} is used as the EP model.
We note that the giant cluster emerges more abruptly at a delayed threshold 
in the case of CDMG (Fig. 1(b)) compared to ER (Fig. 1(a)).

In the current paper, we use CDMG, 
which is an analytically tractable EP model, to understand the
origin of the negative slopes of $H$ at $t_c$ in EP models. 
Using CDMG, we are also able to understand why $\dot{H}(t)$ is minimum at $t=t_c$ and why $\ddot{H}(t)$
diverges at $t=t_c$ in EP models, as reported in~\cite{vieira}.

The rest of this paper is organized as follows.
In Sec.~\ref{sec:intro_dCR}, we briefly introduce the CDMG we study. 
In Sec.~\ref{sec:negativeHprime}, we explore 
why $\dot{H}(t_c)<0$, meaning that $H$ is not maximum at $t_c$
using scaling forms of $n_s$.
In Sec.~\ref{sec:scaling}, we establish the scaling behaviors of $H$, $\dot{H}$, and $\ddot{H}$, and in Sec.~\ref{sec:discussion}, we discuss the potential for the demonstrated theory using CDMG to be applied to various EP models.

\section{Explosive percolation model}
\label{sec:intro_dCR}

In this section, we introduce the CDMG studied in this paper.
In this model, $N$ nodes are isolated $(t = 0)$ at the beginning.
For each link attachment $(t \rightarrow t + 1/N)$, the following process (i)--(ii) is repeated twice to select a pair of nodes to attach the link.
(i) We choose $m$ number of nodes randomly. (ii)   
Among the $m$ randomly chosen nodes, a node belonging to the smallest cluster
is selected.

In this model, as $m$ becomes larger, suppression effects on the growth of large clusters 
increase such that a giant cluster emerges more abruptly but continuously 
at the threshold for finite $m$~\cite{dacosta_prl, dacosta_pre, dacosta_pre_numeric, jan, jan2, grassberger, riordan, cho_science}.
Otherwise, the giant cluster emerges discontinuously at the threshold when $m \rightarrow \infty$~\cite{ep_growing, cho_science, riordan, jan, hybrid}.
We note that CDMG with $m=1$ is equal to the ER model.

The rate equation for $n_s$ in this model is given by
\begin{equation}
\frac{\partial n_s(t)}{\partial t} = \sum_{u+v=s}q_uq_v-2q_s,
\label{Eq:EP_rate}
\end{equation}
where
\begin{equation*}
q_s(t) = \sum_{k=1}^m \binom{m}{k} (sn_s)^k \left[ 1 - \sum_{u=1}^s un_u\right]^{m-k}.
\end{equation*}
Here, $q_s$ is the probability that the size of the smallest cluster among the 
clusters to which $m$ randomly chosen nodes belong is $s$~\cite{dacosta_pre}.
In this paper, we perform simulation or solve Eq.~(\ref{Eq:EP_rate}) numerically to get data for CDMG depending on the situation.

\section{origin of the negative slope of $H$ at $t_c$ in the explosive percolation model}
\label{sec:negativeHprime}

\begin{figure}[t!]
\includegraphics[width=1.0\linewidth]{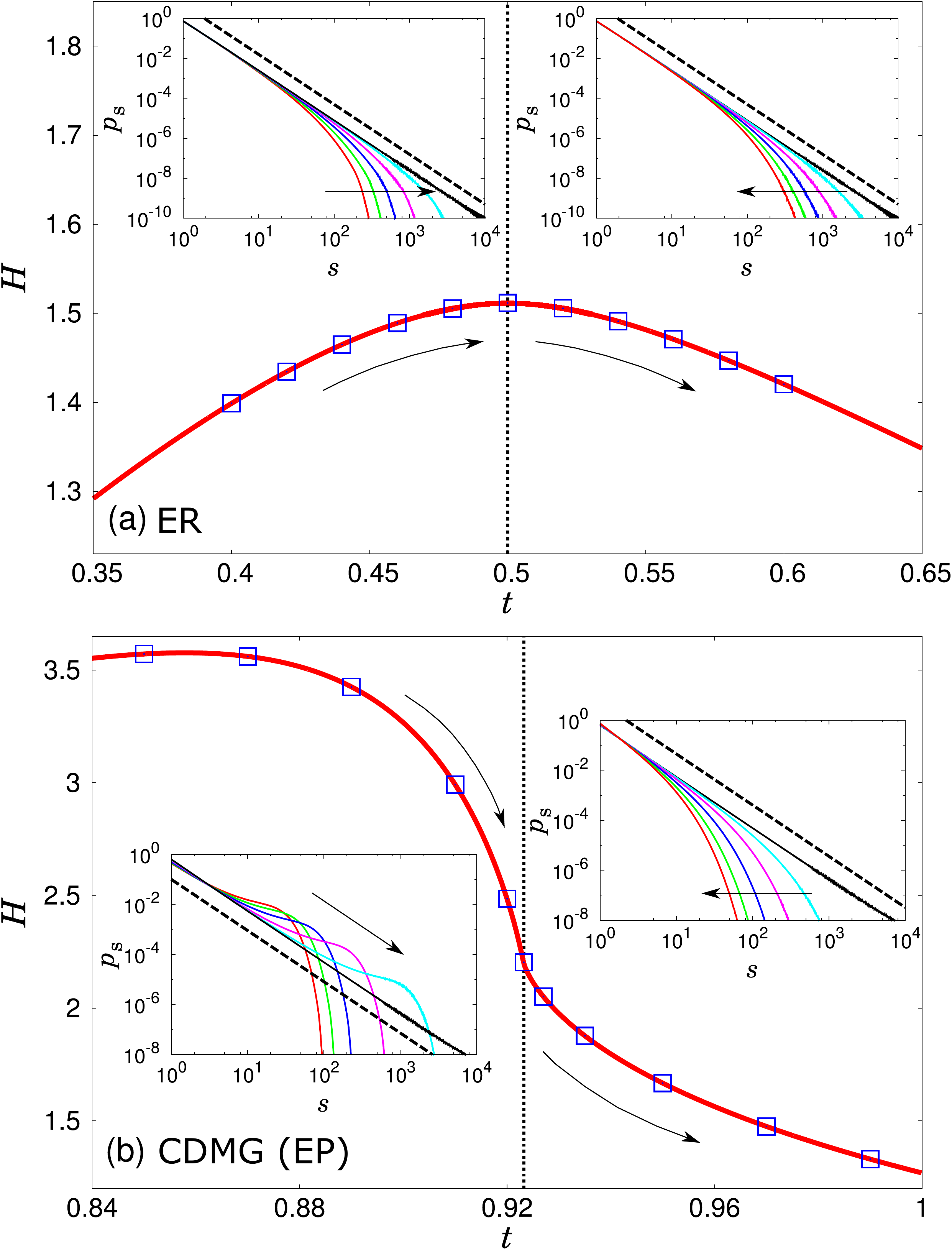}
\caption{
Simulation results with $N = 1.024 \times 10^7$.
(a) $H$ (solid line) vs. $t$ near $t_c=0.5$ (dotted line) in the ER model. 
Left inset: $p_s$ vs. $s$ for $t=0.4$, $0.42$, $0.44$, $0.46$, $0.48$, and $0.5$ from the left.
Right inset: $p_s$ vs. $s$ for $t=0.5$, $0.52$, $0.54$, $0.56$, $0.58$, and $0.6$ from the right. 
In both insets, the slope of the dashed line is $-2.5$. $H(t)$ at the values of $t$ used in both insets are shown in the main panel as $\square$ symbols.
(b) $H$ (solid line) vs. $t$ near $t_c=0.923207508$ (dotted line) in the CDMG model with $m=2$. Left inset: $p_s$ vs. $s$ for $t=0.85$, $0.87$, $0.89$, $0.91$, $0.92$, and $0.923208$ from the left.
Right inset: $p_s$ vs. $s$ for $t=0.923208$, $0.927$, $0.935$, $0.95$, $0.97$, and $0.99$ from the right.
In both insets, the slope of the dashed line is $-2.04762$. $H(t)$ at the values of $t$ used in both insets are shown in the main panel as $\square$ symbols.} 
\label{Fig:ER_EP_psdist}
\end{figure}

We first briefly discuss why $H$ in the CDMG model is decreasing at $t=t_c$ ($\dot{H}(t_c)<0$) 
whereas $H$ in the ER model is maximum at $t=t_c$
through consideration of the distinct scaling behaviors of $p_s$ between the two models.
In the ER model, the scaling form of $n_s$ is given by
\begin{equation}
n_s = \frac{1}{\sqrt{2\pi}}s^{-\tau}{\text {exp}}(-2|t-t_c|^{1/\sigma}s)
\label{Eq:ER_scaling}
\end{equation}
for both $t < t_c$ and $t > t_c$, where $\tau = 5/2$ and $\sigma = 1/2$.
Then, the scaling form of $p_s(t) = A(t)n_s(t) = n_s(t)/(\sum_{s=1}^{\infty}n_s(t))$ with Eq.~(\ref{Eq:ER_scaling}) should satisfy
$p_s(t_c-\delta t)=p_s(t_c+\delta t)$ for $1 \leq s < \infty$.  
Therefore, $H(t)$ calculated using this scaling form of $p_s$ is symmetric with respect to $t=t_c$ such that
it would have a (local) minimum or (local) maximum at $t=t_c$. We check numerically that $H(t)$ calculated using the scaling form of $p_s$ 
is indeed maximum at $t=t_c$ (not shown here).

In CDMG, the scaling form of $n_s$ is given differently for $t < t_c$ and $t > t_c$. Specifically, it is known that the
scaling form of $n_s$ in CDMG is given by
\begin{equation}
n_s = \left\{\begin{array}{lll}
s^{-\tau}f_1((t_c-t)^{1/\sigma}s) & \textrm{~for~} & t \leq t_c \\\\
s^{-\tau}f_2((t-t_c)^{1/\sigma}s) & \textrm{~for~} & t \geq t_c
\end{array}\right.
\label{Eq:EP_ns_scalingform}
\end{equation}
satisfying $f_1(0)=f_2(0)$,
where $\sigma = 1-(2m-1)(\tau-2)$ with $\tau \approx 2 + {\text {exp}}(-1.43 m)$ 
for $m \geq 2$~\cite{dacosta_pre}.
We remark that $(\tau-1)/\sigma-1 > 0$ and $(\tau-1)/\sigma-2 < 0$.

From now on, we use $a_0$ to denote $a_0=f_1(0)=f_2(0)$ in common, and thus $a_0$ is the amplitude of $n_s$ at $t=t_c$, whose value depends on $m$.
Then, the scaling form of $p_s(t) = A(t)n_s(t)=n_s(t)/(\sum_{s=1}^{\infty}n_s(t))$ with Eq.~(\ref{Eq:EP_ns_scalingform}) 
satisfies $p_s(t_c-\delta t) \neq p_s(t_c + \delta t)$ if $\delta t > 0$.
Consequently, $H(t)$ calculated using this scaling form of $p_s$ is not symmetric at $t=t_c$. 
We can briefly understand why the maximum of $H$ does not appear at $t=t_c$ in this way.

In Fig.~\ref{Fig:ER_EP_psdist}, we present $p_s$ and $H$ obtained by simulation 
to support the brief discussion presented above on the different behavior of $H$ in ER and CDMG
based on the different scaling forms of $p_s$ in the two models.
In Fig.~\ref{Fig:ER_EP_psdist}(a), $p_s$ obtained by simulation in the ER model looks almost symmetric
with respect to $t=t_c$, as shown in the insets, 
such that the assumption $p_s(t_c-\delta t)=p_s(t_c+\delta t)$ $(1 \leq s < \infty)$ derived using the
scaling form of $n_s$ (Eq.~(\ref{Eq:ER_scaling})) is reasonable. As expected from this assumption, $H$ obtained by simulation in 
the ER model looks almost symmetric with respect to $t=t_c$, leading to $H$ having its maximum at $t=t_c$.
In Fig.~\ref{Fig:ER_EP_psdist}(b), $p_s$ obtained by simulation in CDMG has a bump in the large $s$ region for $t<t_c$,
whereas it does not have a bump for $t>t_c$, as shown in the insets, evidencing that
$p_s(t_c-\delta t) \neq p_s(t_c+\delta t)$ if $\delta t>0$ as derived
using the scaling form of $n_s$ (Eq.~(\ref{Eq:EP_ns_scalingform})).
Therefore, $H$ obtained by simulation is not symmetric at $t=t_c$ such that it is not maximum at $t=t_c$.

\begin{figure*}[t!]
\includegraphics[width=1.0\linewidth]{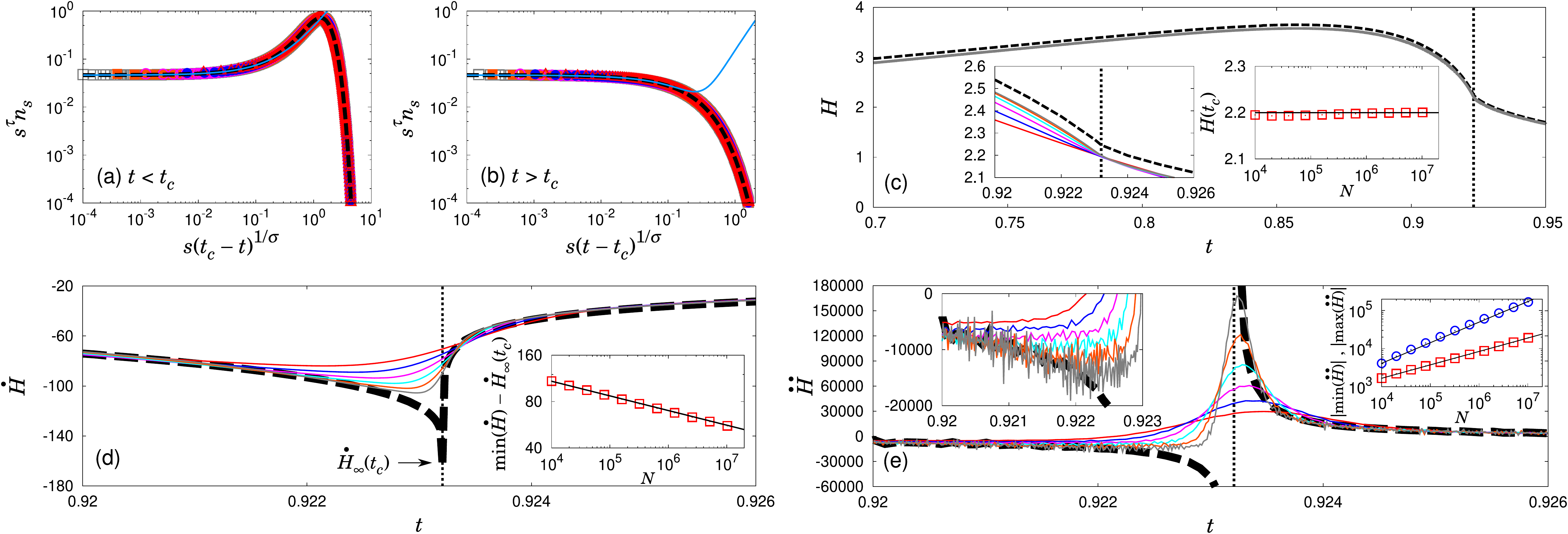}
\caption{
We use the CDMG model with $m = 2$ for this figure. (a) Data collapse of $s^{\tau}n_s$ vs. $s(t_c-t)^{1/\sigma}$
for $t = 0.9$ $(\triangle)$, $0.91$ $(\bullet)$, $0.92$ $(\circ)$, $0.922$ $(\blacksquare)$, and $0.923$ $(\square)$
obtained by solving Eq.~(\ref{Eq:EP_rate}) numerically. 
The dashed line is the scaling function $f_1(x)$ and the solid line is $a_0 + a_1x^{\sigma} + a_2x^{2\sigma}$.
$f_1(x)$ is obtained by connecting the adjacent points of $s^{\tau}n_s$ vs. $s(t_c-t)^{1/\sigma}$ for $t=0.923$ $(\square)$.
(b) Data collapse of $s^{\tau}n_s$ vs. $s(t-t_c)^{1/\sigma}$ 
for $t = 0.9235$ $(\square)$, $0.924$ $(\blacksquare)$, $0.925$ $(\circ)$, $0.926$ $(\bullet)$, and 
$0.927$ $(\triangle)$ obtained by solving Eq.~(\ref{Eq:EP_rate}) numerically. The dashed line is the scaling function
$f_2(x)$ and the solid line is $a_0 - a_1x^{\sigma} + a_2x^{2\sigma}$.
$f_2(x)$ is obtained by connecting the adjacent points of $s^{\tau}n_s$ vs. $s(t-t_c)^{1/\sigma}$ for $t=0.9235$ $(\square)$.
(a, b) We use $\tau = 2.04762$, $\sigma = 0.857$, $t_c = 0.923207508$, $a_0=0.04618$, $a_1=0.165563$, and $a_2=0.27041$.
(c--e) $H$, $\dot{H}$, and $\ddot{H}$ near $t_c$ (dotted line).
(c) The solid line is $H$ obtained from simulation with $N = 1.024 \times 10^7$, and 
the dashed line is $H$ calculated with $p_s$ $(1 \leq s < \infty)$ estimated using the scaling functions 
$f_1(x)$ and $f_2(x)$ for $t \leq t_c$ and $t \geq t_c$, respectively.
Left inset: Solid lines are $H$ obtained from simulation with $N/10^4 = 32$, $64$, $128$, $256$, $512$, and $1024$ from below,
and the dashed line is the same line in the main panel. 
Right inset: $H(t_c)$ obtained by simulation for various values of $N$. $H(t_c)$ seems to saturate at approximately $2.2$ as $N$ increases.
(d) Solid lines are $\dot{H}$ obtained from simulation with $N/10^4 = 32$, $64$, $128$, $256$, $512$, and $1024$ from above,
and the dashed line is $\dot{H}$ for $H$ represented by the dashed line in (c)
with the theoretical value $\dot{H}_{\infty}(t_c)\approx -161.2332$ at $t_c$.
Inset: $\text{min}(\dot{H})$  for $\dot{H}$ obtained from simulation
converges to $\dot{H}_{\infty}(t_c)$ with $N$
as $\text{min}(\dot{H})-\dot{H}_{\infty}(t_c) \sim N^{-0.10}$. The solid line is a guide to the eye.
(e) Solid lines are $\ddot{H}$ obtained from simulation with $N/10^4 = 32$, $64$, $128$, $256$, $512$, and $1024$ from above (below) 
for $t < t_c$ $(t > t_c)$.
Dashed lines are $\ddot{H}$ for $H$ represented by the dashed line in (c).
Left inset: Enlarged plot of the lines for $t < t_c$ in the main panel.
Right inset: $|\text{min}(\ddot{H})|$ $(\square)$ and $|\text{max}(\ddot{H})|$ $(\circ)$ for 
$\ddot{H}$ obtained by simulation diverge with $N$ as $N^{0.35}$ and $N^{0.54}$, respectively. 
Solid lines are guides to the eye.} 
\label{Fig:EP_polylog_H_Hprime}
\end{figure*}

We show that the estimated $\dot{H}(t_c)$ in the $N \rightarrow \infty$ limit denoted by $\dot{H}_{\infty}(t_c)$ is negative and finite.
For this purpose, 
we obtain $\dot{H}_{\infty}(t_c)$ by substituting $t=t_c$ 
after differentiating both sides of Eq.~(\ref{Eq:Hdef}) with respect to $t$ and deleting the term $\sum_{s=1}^{\infty}\dot{p}_s(t)$ 
by the normalization condition $\sum_{s=1}^{\infty}p_s(t)=1$. Then, 
we calculate 
\begin{equation}
\dot{H}_{\infty}(t_c) = -\sum_{s=1}^{\infty}\dot{p}_s(t_c)
{\text {log}}_2p_s(t_c) 
\label{Eq:dotHstar}
\end{equation} 
using the scaling form of $n_s$, where $p_s(t_c)=A(t_c)n_s(t_c)$ and
$\dot{p}_s(t_c)=\dot{A}(t_c)n_s(t_c)+A(t_c)\dot{n}_s(t_c)$ with $A(t_c)=1/(\sum_{s=1}^{\infty}n_s(t_c))$
and $\dot{A}(t_c)=-A(t_c)^2\sum_{s=1}^{\infty}\dot{n}_s(t_c)$.

For small $x=(t_c-t)^{1/\sigma}s \ll 1$ $(x=(t-t_c)^{1/\sigma}s \ll 1)$, 
expansion of the scaling function $f_1(x)$ $(f_2(x))$ up to $\mathcal{O}(x^{2\sigma})$ is given by
$f_1(x) \approx a_0 + a_1x^{\sigma} +a_2x^{2\sigma}$ $(f_2(x) \approx a_0 - a_1x^{\sigma} + a_2x^{2\sigma})$,
where $a_1, a_2$ are analytic functions of $a_0, m, \tau$~\cite{dacosta_pre}.
We check these behaviors of $f_1(x)$ and $f_2(x)$ for small $x$
in Fig.~\ref{Fig:EP_polylog_H_Hprime}(a) and (b), respectively. 
By inserting $x=0$ for $t=t_c$, we can obtain $n_s(t_c)=a_0s^{-\tau}$ and $\dot{n}_s(t_c) = -a_1s^{\sigma - \tau}$
such that 
$A(t_c) = 1/(a_0\sum_{s=1}^{\infty}s^{-\tau})$ and 
$\dot{A}(t_c)=A(t_c)^2a_1\sum_{s=1}^{\infty}s^{\sigma-\tau}$. 
Substituting $p_s(t_c)$ and $\dot{p}_s(t_c)$ expressed using these results in Eq.~(\ref{Eq:dotHstar}),
$\dot{H}_{\infty}(t_c)$ can be written as
\begin{widetext}
\begin{equation}
\dot{H}_{\infty}(t_c)=-\frac{\tau a_1}{a_0(\sum_{s=1}^{\infty}s^{-\tau})^2}\left[
\left(\sum_{u=1}^{\infty} u^{\sigma-\tau}{\text {log}}_2 u\right)\left(\sum_{v=1}^{\infty}v^{-\tau}\right)
-\left(\sum_{u=1}^{\infty}u^{\sigma-\tau}\right)\left(\sum_{v=1}^{\infty}v^{-\tau}{\text {log}}_2v\right)\right].
\label{Eq:Hprimeinfty}
\end{equation}
\end{widetext}
We remark that $\dot{H}_{\infty}(t_c)$ is negative and finite irrespective of $m$ 
because $a_1 > 0$ and $\tau-\sigma > 1$~\cite{dacosta_pre}.

In Fig.~\ref{Fig:EP_polylog_H_Hprime}(c), we calculate $H$ with the $p_s$ $(1 \leq s < \infty)$ estimated using the scaling functions
$f_1(x)$ and $f_2(x)$ obtained numerically in Fig.~\ref{Fig:EP_polylog_H_Hprime}(a) and (b), respectively.
Here, it can be seen that $H(t)$ calculated with this method and $H(t)$ obtained by simulation show similar decreasing curves near $t=t_c$. 
In Fig.~\ref{Fig:EP_polylog_H_Hprime}(d),~$\dot{H}$ obtained by numerically computing the first derivative of $H$
in Fig.~\ref{Fig:EP_polylog_H_Hprime}(c) seems to decrease to $\dot{H}_{\infty}(t_c)$ as $t \rightarrow t_c$.
In particular, we show that the minimum of $\dot{H}(t)$ obtained by simulation decreases to $\dot{H}_{\infty}(t_c)$ with $N$.
These results support that the exact value of $\dot{H}(t_c)$ as $N \rightarrow \infty$ 
is approximately the same as $\dot{H}_{\infty}(t_c)$.
In Fig.~\ref{Fig:EP_polylog_H_Hprime}(e), $\ddot{H}$ obtained by numerically computing the second derivative of $H$
in Fig.~\ref{Fig:EP_polylog_H_Hprime}(c) seems to diverge negatively (positively) at the left (right) limit of the threshold.
To support this expectation, we show that the minimum (maximum) of $\ddot{H}$ obtained by simulation diverges negatively (positively) with $N$.

\section{Scaling behaviors of $H$, $\dot{H}$, and $\ddot{H}$ near $t_c$ in the explosive percolation model}
\label{sec:scaling}

In the previous section, we showed that $H$ decreases at $t_c$
with a negative slope, where the estimated $\dot{H}(t_c)$ in the $N \rightarrow \infty$ limit is 
$\dot{H}_{\infty}(t_c)$ (Eq.~(\ref{Eq:Hprimeinfty})).
$\ddot{H}(t)$ seems to diverge to negative (positive) infinity as $t$ approaches $t_c$ from the left (right).
In this section, we analyze the scaling behaviors of $H(t)-H(t_c)$, $\dot{H}(t)-\dot{H}_{\infty}(t_c)$, and
$\ddot{H}(t)$ as $|t-t_c| \rightarrow 0$ for $t < t_c$ and $t > t_c$ separately.

For ease of analysis, $H$ is simplified by using an approximation that holds for $|t-t_c| \ll 1$.
When a link is attached, it connects two nodes belonging to either distinct clusters or the same cluster.
Up to the emergence of the giant cluster $(\text{for}~t \leq t_c)$, the first event among the two events occurs dominantly
at each link attachment. Therefore, $\sum_{s=1}^{\infty} n_s$ decreases by $1/N$
as $t \rightarrow t+1/N$ up to $t=t_c$, such that
$\sum_{s=1}^{\infty} n_s = 1-t$ holds for $t \leq t_c$. 
This allows us to use the approximation $\sum_{s=1}^{\infty}n_s \approx 1-t$, 
which is equal to $A(t) \approx 1/(1-t)$ for $|t-t_c| \ll 1$
where the scaling behaviors are studied. Applying this $H$ approximation, 
Eq.~(\ref{Eq:Hdef}) with $p_s = An_s$ becomes
\begin{equation}
H = {\text {log}}_2(1-t)+\frac{1}{(1-t)}\left[-\int_{1}^{\infty}n_s {\text {log}}_2 n_sds\right]
\label{Eq:Happrox}
\end{equation}
for $|t-t_c| \ll 1$ after being approximated by $\sum_{s=1}^{\infty}n_s {\text {log}}_2n_s \approx 
\int_1^{\infty}n_s {\text{log}}_2n_sds$,
where the normalization condition $\sum_{s=1}^{\infty}An_s=1$ is used consistently.

We can then also obtain an approximated $\dot{H}$ as
\begin{eqnarray}
\dot{H} &=& -\frac{1}{\text{ln}2}\bigg[\frac{1}{(1-t)} + \frac{1}{(1-t)^2}
\int_1^{\infty}n_s\text{ln}n_sds
\notag \\
&+&\frac{1}{(1-t)}\bigg\{\int_1^{\infty}\dot{n}_s{\text{ln}}n_sds 
+ \int_1^{\infty}\dot{n}_sds\bigg\}
\bigg]
\label{Eq:Hdotapprox}
\end{eqnarray}
by differentiating $H(t)$ in Eq.~(\ref{Eq:Happrox}) 
with respect to $t$.

\subsection{Below the threshold $t < t_c$}
\begin{figure}[t!]
\includegraphics[width=1.0\linewidth]{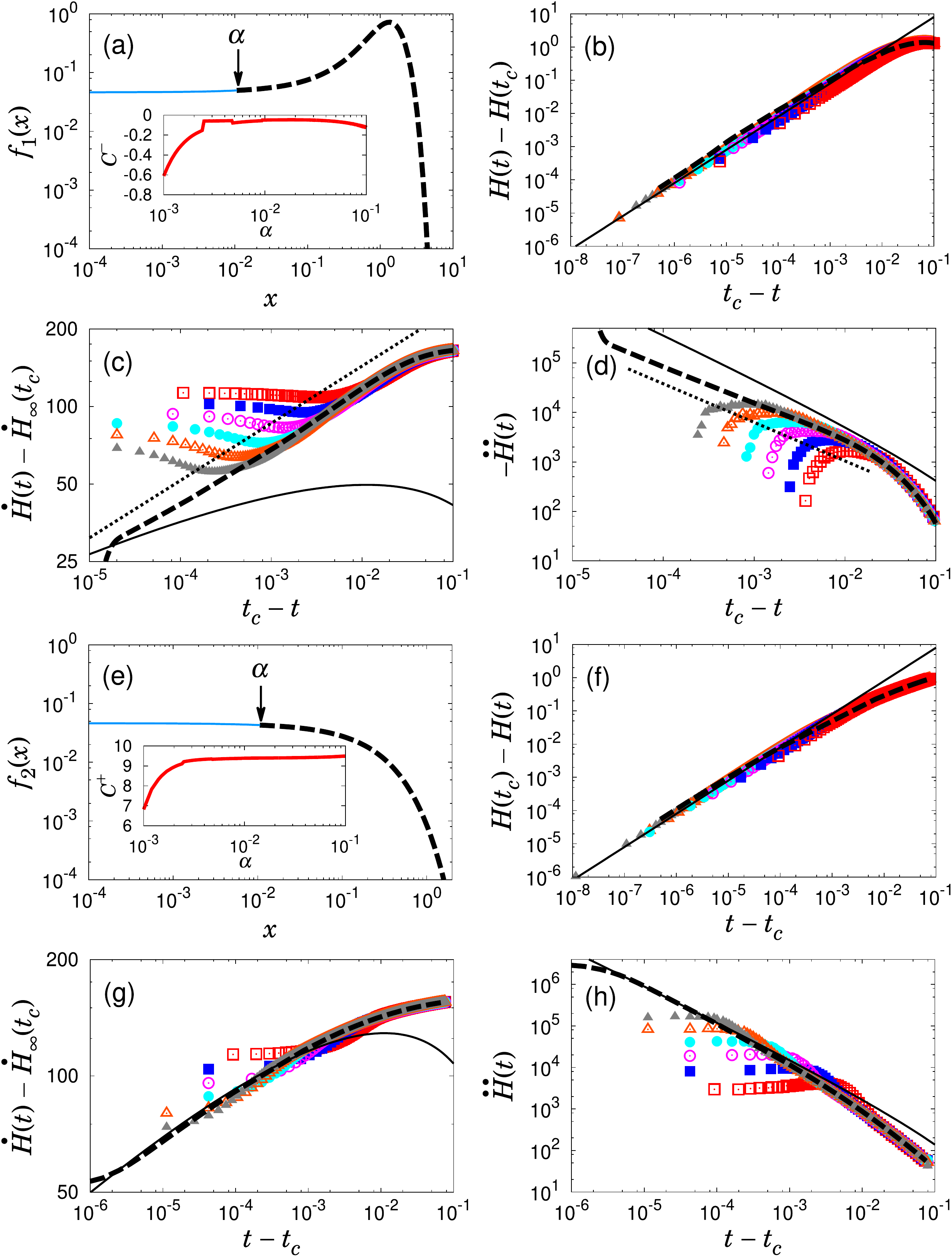}
\caption{
Scaling behaviors of the CDMG model with $m=2$ for (a--d) $t < t_c$ and (e--h) $t > t_c$.
(a--h) We use $\tau = 2.04762$, $\sigma = 0.857$, $a_0=0.04618$, $a_1=0.165563$, and $a_2=0.27041$.
(a) For the $f_1(x)$ obtained in Fig.~\ref{Fig:EP_polylog_H_Hprime}(a), if we use the approximation 
$f_1(x) \approx a_0 + a_1x^{\sigma} + a_2x^{2\sigma}$ for $x < \alpha$, 
the calculated $C^-$ is negative, independent of $\alpha$, as shown in the inset.
(b) $H(t) - H(t_c)$ obtained from simulation with $N/10^4=1$ $(\square)$, $4$ $(\blacksquare)$, 
$16$ $(\circ)$, $64$ $(\bullet)$, $256$ $(\triangle)$, and $1024$ $(\blacktriangle)$, as well as from 
solving Eq.~(\ref{Eq:EP_rate}) numerically (dashed line).
The slope of the solid line is $1$.
(c) The symbols and the dashed line are $\dot{H}(t) - \dot{H}_{\infty}(t_c)$ for $H(t)$ in (b).
The slope of the dotted line is $(\tau-1)/\sigma-1$ and the solid line is $\propto -(t_c-t)^{(\tau-1)/\sigma-1}{\text {log}}_2(t_c-t)$.
(d) The symbols and the dashed line are $-\ddot{H}(t)$ for $H(t)$ in (b). 
The slope of the dotted line is $(\tau-1)/\sigma-2$ and the solid line is $\propto -(t_c-t)^{(\tau-1)/\sigma-2}{\text {log}}_2(t_c-t)$.
(e) For the $f_2(x)$ obtained in Fig.~\ref{Fig:EP_polylog_H_Hprime}(b), if we use the approximation 
$f_2(x) \approx a_0 - a_1x^{\sigma} + a_2x^{2\sigma}$ for $x < \alpha$, 
the calculated $C^+$ is positive, independent of $\alpha$ (inset).
(f) $H(t_c)-H(t)$ obtained from simulation with $N/10^4=1$ $(\square)$, $4$ $(\blacksquare)$, 
$16$ $(\circ)$, $64$ $(\bullet)$, $256$ $(\triangle)$, and $1024$ $(\blacktriangle)$, as well as from solving Eq.~(\ref{Eq:EP_rate}) numerically (dashed line).
The solid line is the same as that in (b).
(g) The symbols and the dashed line are $\dot{H}(t) - \dot{H}_{\infty}(t_c)$ for $H(t)$ in (f).
The solid line is $\propto -(t-t_c)^{(\tau-1)/\sigma-1}{\text {log}}_2(t-t_c)$.
(h) The symbols and the dashed line are $\ddot{H}(t)$ for $H(t)$ in (f). 
The solid line is $\propto -(t-t_c)^{(\tau-1)/\sigma-2}{\text {log}}_2(t-t_c)$.} 
\label{Fig:EP_H_Hprime_scaling}
\end{figure}

For $t < t_c$, we use $n_s = (t_c-t)^{\tau/\sigma}x^{-\tau}f_1(x)$ with $x = (t_c-t)^{1/\sigma}s$ such that
$H$ in Eq.~(\ref{Eq:Happrox}) can be written as
\begin{eqnarray}
H &=& {\text {log}}_2(1-t_c+\epsilon^{\sigma})
\notag \\
&+&\frac{1}{(1-t_c+\epsilon^{\sigma})}
\bigg[
-\epsilon^{(\tau-1)}\text{log}_2\epsilon^\tau\int_{\epsilon}^{\infty}x^{-\tau}f_1(x)dx
\notag \\
&-&\epsilon^{(\tau-1)}\int_{\epsilon}^{\infty}x^{-\tau}f_1(x){\text {log}}_2(x^{-\tau}f_1(x))dx
\bigg],
\label{Eq:Happroxfinal}
\end{eqnarray}
where $\epsilon=(t_c-t)^{1/\sigma} \ll 1$.
We divide the interval of integration $[\epsilon, \infty]$ of Eq.~(\ref{Eq:Happroxfinal}) into
two intervals, $[\epsilon, \alpha]$ and $[\alpha, \infty]$ for some $\alpha \ll 1$, 
and use the approximation $f_1(x) \approx a_0 + a_1x^{\sigma} + a_2x^{2\sigma}$ 
for the first interval $[\epsilon, \alpha]$  
(see Sec.~\ref{sec:negativeHprime} and Fig.~\ref{Fig:EP_H_Hprime_scaling}(a)).
Then, the expansion of $H$ up to $\mathcal{O}(\epsilon^{\sigma})=\mathcal{O}(t_c-t)$ is given by 
\begin{eqnarray}
H &\approx& \frac{1}{\text{ln}2}\left[{\text{ln}}(1-t_c) + \frac{\tau a_0}{(1-t_c)(\tau-1)^2} 
- \frac{a_0{\text {ln}}a_0}{(1-t_c)(\tau-1)}\right]
\notag \\
&+&\frac{1}{{\text {ln}}2}\bigg[\frac{1}{(1-t_c)}+\frac{\tau a_1}{(1-t_c)(\tau-\sigma-1)^2}
\notag \\
&-& \frac{a_1(1+{\text {ln}}a_0)}{(1-t_c)(\tau-\sigma-1)} 
- \frac{\tau a_0}{(1-t_c)^2(\tau-1)^2}
\notag \\
&+&\frac{a_0{\text {ln}}a_0}{(1-t_c)^2(\tau-1)}\bigg](t_c-t),
\label{Eq:Hscaling}
\end{eqnarray}
by using expansions
\begin{equation*}
\int_{\epsilon}^{\infty}x^{-\tau}f_1(x)dx \approx 
\frac{a_0}{(\tau-1)}\epsilon^{1-\tau}+\frac{a_1}{(\tau-\sigma-1)}\epsilon^{1+\sigma-\tau}
\end{equation*}
and
\begin{equation*}
\int_{\epsilon}^{\infty}x^{-\tau}f_1(x){\text {log}}_2(x^{-\tau}f_1(x))dx
\end{equation*}
\begin{eqnarray*}
&\approx& \frac{1}{{\text {ln}}2}\bigg[-\frac{\tau a_0}{(\tau-1)}\epsilon^{1-\tau}{\text {ln}}\epsilon
+\bigg\{\frac{a_0{\text {ln}}a_0}{(\tau-1)} - \frac{\tau a_0}{(\tau-1)^2}\bigg\}\epsilon^{1-\tau}
\notag \\
&-&\frac{a_1\tau}{(\tau-\sigma-1)}\epsilon^{1+\sigma-\tau}{\text {ln}}\epsilon
\notag \\
&+&\bigg\{\frac{a_1(1+\text{ln}a_0)}{(\tau-\sigma-1)}-\frac{a_1\tau}{(\tau-\sigma-1)^2}\bigg\}
\epsilon^{1+\sigma-\tau}\bigg].
\end{eqnarray*}
We find that $H(t)-H(t_c) \propto (t_c-t)$ as $t \rightarrow t_c^-$ in Eq.~(\ref{Eq:Hscaling}).

Similarly for $\dot{H}$, we can derive the scaling behavior of $\dot{H}$ as
\begin{eqnarray}
\dot{H} &\approx& 
\frac{1}{\text{ln}2}\bigg[-\frac{1}{(1-t_c)}-\frac{a_0{\text {ln}}a_0}{(1-t_c)^2(\tau-1)}
\notag \\
&+&\frac{\tau a_0}{(1-t_c)^2(\tau-1)^2}
+\frac{a_1(1+\text{ln}a_0)}{(1-t_c)(\tau-\sigma-1)}
\notag \\
&-& \frac{\tau a_1}{(1-t_c)(\tau-\sigma-1)^2}\bigg]
\notag \\
&+&C^-\left[\frac{\sigma}{(\tau-\sigma-1)}\right](t_c-t)^{\frac{(\tau-1)}{\sigma}-1}\text{log}_2(t_c-t),
\notag \\
\label{Eq:Hdotscaling}
\end{eqnarray}
with $(\tau-1)/\sigma-1 > 0$ (see Sec.~\ref{sec:negativeHprime}) by
using $\dot{n}_s = -(t_c-t)^{\tau/\sigma-1}x^{1-\tau}f'_1(x)/\sigma$ 
and approximations $f_1(x) \approx a_0 + a_1x^{\sigma}+a_2x^{2\sigma}$ and
$f'_1(x) \approx \sigma a_1 x^{\sigma-1} + 2\sigma a_2 x^{2\sigma-1}$
in $\epsilon \leq x \leq \alpha$ for some $\alpha \ll 1$ in Eq.~(\ref{Eq:Hdotapprox}),
where $f'_1(x)$ is the differentiation of $f_1(x)$ with respect to $x$ and
\begin{eqnarray}
C^-&=& \frac{\tau}{(1-t_c)\sigma^2}\left(\frac{(\tau-1)}{\sigma}-1\right)
\bigg[-\frac{a_1\sigma\alpha^{1+\sigma-\tau}}{(\tau-\sigma-1)}
\notag \\
&-& \frac{2a_2\sigma\alpha^{1+2\sigma-\tau}}{(\tau-2\sigma-1)}-\alpha^{1-\tau}f_1(\alpha)
\notag \\
&+&(\tau-1)\int_{\alpha}^{\infty}x^{-\tau}f_1(x)dx\bigg].
\label{Eq:Cminus}
\end{eqnarray}
From this result derived from approximation,
we expect the scaling behavior 
$\dot{H}(t)-\dot{H}_{\infty}(t_c) \propto (t_c-t)^{(\tau-1)/\sigma-1}\text{log}_2(t_c-t)$
as $t \rightarrow t_c^-$.
Detailed derivation of Eq.~(\ref{Eq:Hdotscaling}) is given in Appendix~A.


Finally, we obtain the scaling behavior of $\ddot{H}$ by differentiating Eq.~(\ref{Eq:Hdotscaling}) with respect to $t$
such that
\begin{equation}
\ddot{H} \approx -C^{-}(t_c-t)^{(\tau-1)/\sigma-2}\text{log}_2(t_c-t)
\label{Eq:Hddot_tcminus}
\end{equation}
as $t \rightarrow t_c^-$, where $(\tau-1)/\sigma-2 < 0$ (see Sec.~\ref{sec:negativeHprime}).
In Fig.~\ref{Fig:EP_H_Hprime_scaling}(a), $C^-$ calculated numerically using Eq.~(\ref{Eq:Cminus})
has negative values regardless of $\alpha \ll 1$.
For this reason, as shown in Fig.~\ref{Fig:EP_polylog_H_Hprime}(e), $\ddot{H}$
seems to diverge negatively as $t \rightarrow t_c$
from the left. 
We note that the exact value of $C^-$ without the approximation using $\alpha \ll 1$
is obtained by substituting the closed form of $f_1(x)$ into Eq.~(\ref{Eq:Cminus}) and 
taking the limit as $\alpha \rightarrow 0$.
Moreover, the derivation of the scaling behaviors in Eqs.~(\ref{Eq:Hscaling})--(\ref{Eq:Hddot_tcminus}) considers 
the expansion of the scaling function $f_1(x)$ up to the $x^{2\sigma}$ term for small $x$
to reflect the bump of $n_s$ appearing at $t < t_c$ as shown in Fig.~\ref{Fig:EP_H_Hprime_scaling}(a).

In Fig.~\ref{Fig:EP_H_Hprime_scaling}(b), we find that the scaling behavior of $H(t)-H(t_c)$ fits well with the theory $H(t)-H(t_c) \propto (t_c-t)$.
In Fig.~\ref{Fig:EP_H_Hprime_scaling}(c) and (d), we check the scaling behaviors of
$\dot{H}(t)-\dot{H}_{\infty}(t_c)$ and $\ddot{H}(t)$
as $t \rightarrow t_c^-$ using the data (dashed lines and symbols).
Here, the data for $\dot{H}(t)-\dot{H}_{\infty}(t_c)$ and $\ddot{H}(t)$ 
seem to fit better with $(t_c-t)^{(\tau-1)/\sigma-1}$ and $(t_c-t)^{(\tau-1)/\sigma-2}$
than with the theoretical curves $(t_c-t)^{(\tau-1)/\sigma-1}{\text {log}}_2(t_c-t)$ and $(t_c-t)^{(\tau-1)/\sigma-2}{\text {log}}_2(t_c-t)$, respectively.
To resolve these discrepancies, we obtain that the next dominant term of $\dot{H}(t)-\dot{H}_{\infty}(t_c)$
for $(t_c-t) \rightarrow 0$ is $\mathcal{O}((t_c-t)^{(\tau-1)/\sigma-1})$,
where the coefficient of this term is large enough such that the term is dominant
in the range of $(t_c-t)$ in Fig.~\ref{Fig:EP_H_Hprime_scaling}(c).
For similar reasons, the term $\mathcal{O}((t_c-t)^{(\tau-1)/\sigma-2})$
is dominant for $\ddot{H}(t)$ in the range of $(t_c-t)$ in Fig.~\ref{Fig:EP_H_Hprime_scaling}(d).
Therefore, the discrepancy occurs because $(t_c-t)$ in Fig.~\ref{Fig:EP_H_Hprime_scaling}(c) and (d)
is not small enough to reflect the $(t_c-t) \rightarrow 0$ limit used to derive the theoretical curves. 
Details of this discussion are provided in Appendix B.

\subsection{Above the threshold $t > t_c$}
\label{sec:scaling_supercritical}
For $t > t_c$, we use the scaling form $n_s = (t-t_c)^{\tau/\sigma}x^{-\tau}f_2(x)$ for $x = (t-t_c)^{1/\sigma}s$
with the approximations $f_2(x) \approx a_0 - a_1x^{\sigma}+a_2x^{2\sigma}$ 
and $f'_2(x) \approx -\sigma a_1 x^{\sigma-1} + 2\sigma a_2x^{2\sigma-1}$
in the interval $\epsilon \leq x \leq \alpha$
for some $\alpha \ll 1$, where $\epsilon = (t-t_c)^{1/\sigma}$
(see Sec.~\ref{sec:negativeHprime} and Fig.~\ref{Fig:EP_H_Hprime_scaling}(e)).
For $t > t_c$, derivations of the scaling behaviors of $H(t)-H(t_c)$, $\dot{H}(t)-\dot{H}_{\infty}(t_c)$,
and $\ddot{H}(t)$ are similar to those for $t < t_c$; therefore, we present here only the results of these scaling behaviors.

At first, the expansion of $H$ up to $\mathcal{O}(\epsilon^{\sigma})=\mathcal{O}(t-t_c)$ is equal to Eq.~(\ref{Eq:Hscaling}),
such that $H(t)-H(t_c) \propto -(t-t_c)$ as $t \rightarrow t_c^+$.
Next, the scaling behavior of $\dot{H}$ is derived as
\begin{eqnarray}
\dot{H} &\approx& 
\frac{1}{\text{ln}2}\bigg[-\frac{1}{(1-t_c)}-\frac{a_0{\text {ln}}a_0}{(1-t_c)^2(\tau-1)}
\notag \\
&+&\frac{\tau a_0}{(1-t_c)^2(\tau-1)^2}
+\frac{a_1(1+\text{ln}a_0)}{(1-t_c)(\tau-\sigma-1)}
\notag \\
&-& \frac{\tau a_1}{(1-t_c)(\tau-\sigma-1)^2}\bigg]
\notag \\
&-&C^+\left[\frac{\sigma}{(\tau-\sigma-1)}\right](t-t_c)^{\frac{(\tau-1)}{\sigma}-1}\text{log}_2(t-t_c),
\notag \\
\label{Eq:Hdotscalingtcplus}
\end{eqnarray}
where 
\begin{eqnarray}
C^+&=& \frac{\tau}{(1-t_c)\sigma^2}\left(\frac{(\tau-1)}{\sigma}-1\right)
\bigg[\frac{a_1\sigma\alpha^{1+\sigma-\tau}}{(\tau-\sigma-1)}
\notag \\
&-& \frac{2a_2\sigma\alpha^{1+2\sigma-\tau}}{(\tau-2\sigma-1)}-\alpha^{1-\tau}f_2(\alpha)
\notag \\
&+&(\tau-1)\int_{\alpha}^{\infty}x^{-\tau}f_2(x)dx\bigg].
\label{Eq:Cplus}
\end{eqnarray}
Therefore, we expect the scaling behavior to be
$\dot{H}(t)-\dot{H}_{\infty}(t_c) \propto (t-t_c)^{(\tau-1)/\sigma-1}{\text {log}}_2(t-t_c)$
as $t \rightarrow t_c^+$.
Finally, the scaling behavior of $\ddot{H}$ obtained 
by differentiating Eq.~(\ref{Eq:Hdotscalingtcplus}) is
\begin{equation}
\ddot{H} \approx -C^+(t-t_c)^{\frac{(\tau-1)}{\sigma}-2}\text{log}_2(t-t_c) 
\end{equation}
as $t \rightarrow t_c^+$. 
In Fig.~\ref{Fig:EP_H_Hprime_scaling}(e), 
$C^+$ calculated numerically using Eq.~(\ref{Eq:Cplus}) has positive values
regardless of $\alpha \ll 1$.
For this reason, as seen in Fig.~\ref{Fig:EP_polylog_H_Hprime}(e),
$\ddot{H}$ seems to diverge positively as $t \rightarrow t_c$ from the right.
We note that the exact value of $C^+$ without the approximation using $\alpha \ll 1$
is obtained by substituting the closed form of $f_2(x)$ into Eq.~(\ref{Eq:Cplus}) and 
taking the limit as $\alpha \rightarrow 0$.

In Fig.~\ref{Fig:EP_H_Hprime_scaling}(f), we find that the scaling behavior of $H(t)-H(t_c)$ 
fits well with the theory $H(t)-H(t_c) \propto -(t-t_c)$.
Next, we plot the data for $\dot{H}(t)-\dot{H}_{\infty}(t_c)$ and $\ddot{H}(t)=|\ddot{H}(t)|$
in Fig.~\ref{Fig:EP_H_Hprime_scaling}(g) and (h), respectively.
Unlike the $t < t_c$ case, here the data fit well with the theoretical curves 
$(t-t_c)^{(\tau-1)/\sigma-1}\text{log}_2(t-t_c)$ for $\dot{H}(t)-\dot{H}_{\infty}(t_c)$
and $(t-t_c)^{(\tau-1)/\sigma-2}\text{log}_2(t-t_c)$ for $\ddot{H}(t)$.
To understand this difference, 
we first derived that the next dominant terms for $\dot{H}(t)-\dot{H}_{\infty}(t_c)$ and $\ddot{H}(t)$ 
are $\mathcal{O}((t-t_c)^{(\tau-1)/\sigma-1})$ and $\mathcal{O}((t-t_c)^{(\tau-1)/\sigma-2})$, respectively, even for $t>t_c$. 
Then we confirmed in both cases that
the coefficients of the dominant and the next dominant terms are comparable 
to each other, unlike the $t < t_c$ case, such that $(t-t_c)$ 
in Fig.~\ref{Fig:EP_H_Hprime_scaling}(g) and (h)
is small enough to reflect the $(t-t_c)\rightarrow 0$ limit used to derive the theoretical curves.

\section{Discussion}
\label{sec:discussion}

\begin{figure}[t!]
\includegraphics[width=1.0\linewidth]{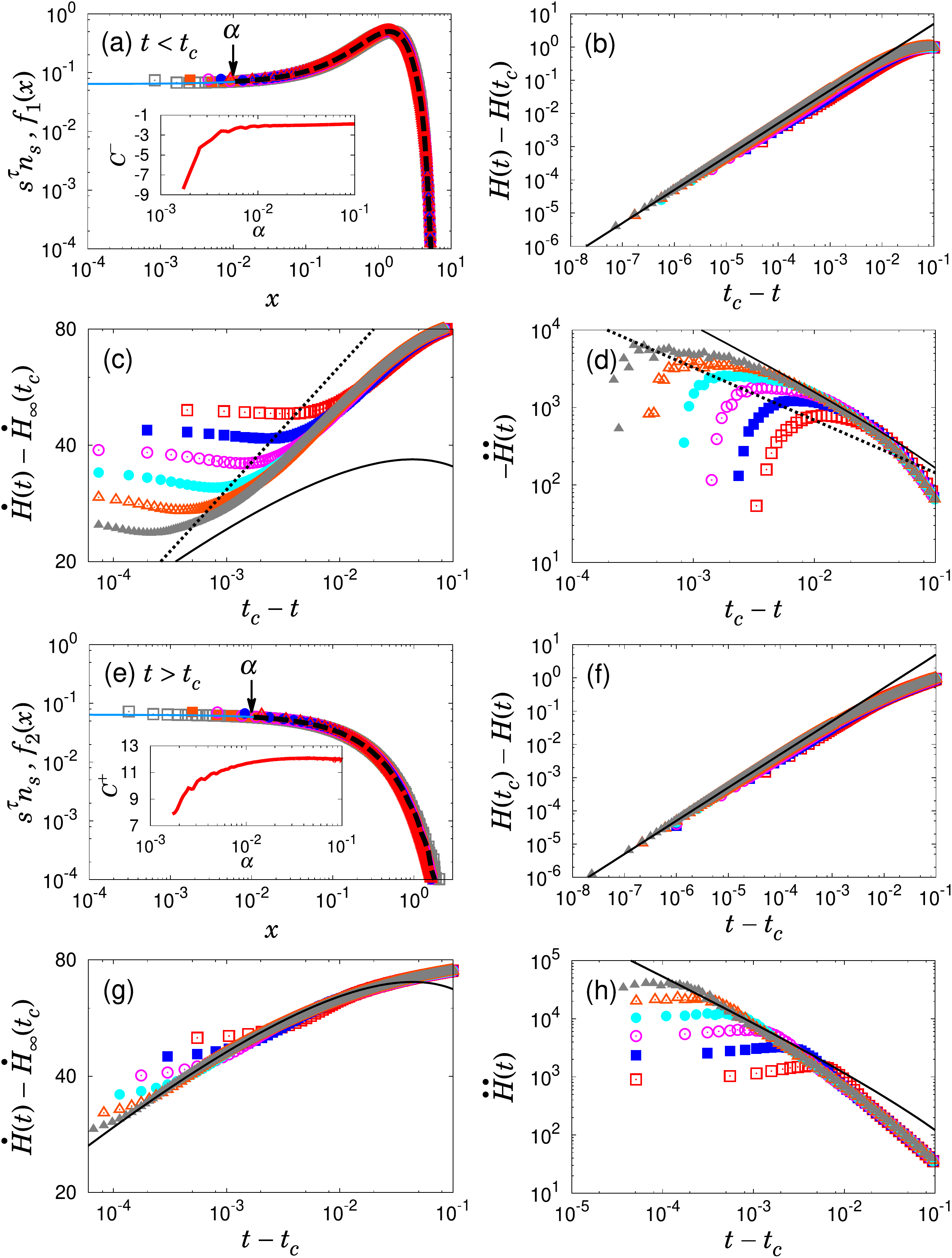}
\caption{
Scaling behaviors of the product rule for (a--d) $t < t_c$ and (e--h) $t > t_c$, where $t_c = 0.888449$.
(a--h) We use $\tau = 2.06$, $\sigma = 0.8$, $a_0=0.064$, $a_1=0.22$, and $a_2=0.25$.
(a) Data collapse of $s^{\tau}n_s$ vs. $x$ obtained from simulation with $N = 1.024 \times 10^7$
for $t = 0.865$ $(\triangle)$, $0.87$ $(\bullet)$, $0.875$ $(\circ)$, $0.88$ $(\blacksquare)$, and $0.885$ $(\square)$.
The dashed line is $f_1(x)$ for $x > \alpha$ obtained by connecting the adjacent points of $s^{\tau}n_s$ vs. $s(t_c-t)^{1/\sigma}$ for $t=0.885$ $(\square)$.
The solid line is $f_1(x)$ for $x < \alpha$ approximated by $f_1(x) \approx a_0+a_1x^{\sigma}+a_2x^{2\sigma}$.
Using this $f_1(x)$, 
the calculated $C^-$ is negative, independent of $\alpha$, as shown in the inset.
(b) $H(t) - H(t_c)$ obtained from simulation using $N/10^4=1$ $(\square)$, $4$ $(\blacksquare)$, 
$16$ $(\circ)$, $64$ $(\bullet)$, $256$ $(\triangle)$, and $1024$ $(\blacktriangle)$.
The slope of the solid line is $1$.
(c) The symbols are $\dot{H}(t) - \dot{H}_{\infty}(t_c)$ for $H(t)$ in (b).
The slope of the dotted line is $(\tau-1)/\sigma-1$ and the solid line is $\propto -(t_c-t)^{(\tau-1)/\sigma-1}{\text {log}}_2(t_c-t)$.
(d) The symbols are $-\ddot{H}(t)=|\ddot{H}(t)|$ for $H(t)$ in (b). 
The slope of the dotted line is $(\tau-1)/\sigma-2$ and the solid line is $\propto -(t_c-t)^{(\tau-1)/\sigma-2}{\text {log}}_2(t_c-t)$.
(e) Data collapse of $s^{\tau}n_s$ vs. $x$ obtained from simulation with $N = 1.024 \times 10^7$
for $t = 0.92$ $(\triangle)$, $0.91$ $(\bullet)$, $0.9$ $(\circ)$, $0.895$ $(\blacksquare)$, and $0.89$ $(\square)$.
The dashed line is $f_2(x)$ for $x > \alpha$ obtained by connecting the adjacent points of 
$s^{\tau}n_s$ vs. $s(t-t_c)^{1/\sigma}$ for $t=0.89$ $(\square)$. 
Using $f_2(x)$ approximated by $f_2(x) \approx a_0 - a_1x^{\sigma} + a_2x^{2\sigma}$ for $x < \alpha$, 
the calculated $C^+$ is positive, independent of $\alpha$ (inset).
(f) $H(t_c) - H(t)$ obtained from simulation using $N/10^4=1$ $(\square)$, $4$ $(\blacksquare)$, 
$16$ $(\circ)$, $64$ $(\bullet)$, $256$ $(\triangle)$, and $1024$ $(\blacktriangle)$.
The solid line is the same as that in (b).
(g) The symbols are $\dot{H}(t) - \dot{H}_{\infty}(t_c)$ for $H(t)$ in (f).
The solid line is $\propto -(t-t_c)^{(\tau-1)/\sigma-1}{\text {log}}_2(t-t_c)$.
(h) The symbols are $\ddot{H}(t)$ for $H(t)$ in (f). 
The solid line is $\propto -(t-t_c)^{(\tau-1)/\sigma-2}{\text {log}}_2(t-t_c)$.} 
\label{Fig:PR_H_Hprime_scaling}
\end{figure}

In summary, we use the scaling form of $n_s$ in the CDMG model given differently for $t < t_c$ and $t > t_c$
as shown in Eq.~(\ref{Eq:EP_ns_scalingform}), where the scaling functions $f_1(x)$ for $t<t_c$
and $f_2(x)$ for $t>t_c$
are approximated by 
$f_1(x) \approx a_0 + a_1x^{\sigma}+a_2x^{2\sigma}$ and $f_2(x) \approx a_0 - a_1x^{\sigma}+a_2x^{2\sigma}$
for sufficiently small $x \ll 1$.
Then, we obtain the scaling behaviors
$H(t)-H(t_c) \propto -(t-t_c)$, 
$\dot{H}(t)-\dot{H}_{\infty}(t_c) \propto -|t-t_c|^{(\tau-1)/\sigma-1}
\text{log}_2|t-t_c|$, and $|\ddot{H}(t)| \propto |t-t_c|^{(\tau-1)/\sigma-2}{\text{log}}_2|t-t_c|$
as $t \rightarrow t_c$ from both sides,
where $\dot{H}_{\infty}(t_c) < 0$ is the estimated value of $\dot{H}(t)$ at $t=t_c$.
As a result, we find that $H(t)$ decreases at $t=t_c$ $(\dot{H}(t_c) < 0)$, $\dot{H}(t)$ is minimum at $t=t_c$,
and $\ddot{H}(t)$ diverges 
as expected using the simulation in~\cite{vieira}.

If $n_s$ follows the scaling form in  Eq.~(\ref{Eq:EP_ns_scalingform}) with approximations
$f_1(x) \approx a_0+a_1x^{\sigma}+a_2x^{2\sigma}$ and
$f_2(x) \approx a_0-a_1x^{\sigma}+a_2x^{2\sigma}$ 
for sufficiently small $x \ll 1$ with $a_1>0$ and $a_2>0$, then $n_s(t)$ has a bump at the large $s$ region for $t < t_c$,
with the bump disappearing as $t$ exceeds $t_c$.
It is known that the existence of such a bump for $t<t_c$ is a general property of EP models; therefore, we expect that the derivation of the scaling behaviors in CDMG 
discussed in this paper could be applied to general EP models.
To demonstrate, we apply it to the product rule~\cite{science}, the first discovered EP model.
In this model, we perform simulation to obtain $n_s$ for different values of $t$ near $t_c$, 
and estimate $f_1(x)$ and $f_2(x)$ using a data collapse of $s^{\tau}n_s$ vs. $x \equiv s|t-t_c|^{1/\sigma}$
with different values of $t$
for $t < t_c$ and $t > t_c$, respectively.
We assume that $f_1(x) \approx a_0+a_1x^{\sigma}+a_2x^{2\sigma}$ and
$f_2(x) \approx a_0-a_1x^{\sigma}+a_2x^{2\sigma}$ 
for sufficiently small $x \ll 1$ in the product rule as well.
To estimate $a_0$, $a_1$, and $a_2$, we first take the values of $\tau$ and $\sigma$ reported in~\cite{hklee}, and then we find $a_1$ by using $f_1(x)-f_2(x) \approx 2a_1x^{\sigma}$ in the range of small $x$.
Finally, we change the value of $a_0$ and adopt the values of $a_0$ and $a_2$ when 
$f_1(x)+f_2(x)-2a_0 \approx 2a_2x^{2\sigma}$ shows
power-law behavior with a slope of $2\sigma$. With the $f_1(x)$, $f_2(x)$, $a_1$, and $a_2$ obtained in this manner, 
we can predict the scaling behaviors in the product rule similar to those in CDMG, as shown in Fig.~\ref{Fig:PR_H_Hprime_scaling}.

For $t < t_c$, as shown in Fig.~\ref{Fig:PR_H_Hprime_scaling}(a), 
$f_1(x) \approx a_0 + a_1x^{\sigma}+a_2x^{2\sigma}$ using the estimated $a_0$, $a_1$, and $a_2$ 
is reasonable for small $x$, and $C^-$ calculated numerically using Eq.~(\ref{Eq:Cminus}) has negative values regardless of $\alpha \ll 1$.  
Then, the scaling behaviors of $H(t)-H(t_c)$, $\dot{H}(t)-\dot{H}_{\infty}(t_c)$, and $\ddot{H}(t)$ as $t \rightarrow t_c^-$
are checked using the data as shown in Fig.~\ref{Fig:PR_H_Hprime_scaling}(b), (c), and (d), respectively.
In Fig.~\ref{Fig:PR_H_Hprime_scaling}(b), the scaling behavior of $H(t)-H(t_c)$ fits well with the theory $H(t)-H(t_c) \propto (t_c-t)$.
In Fig.~\ref{Fig:PR_H_Hprime_scaling}(c) and (d), the data for $\dot{H}(t)-\dot{H}_{\infty}(t_c)$ and $\ddot{H}(t)$ 
seem to fit better with $(t_c-t)^{(\tau-1)/\sigma-1}$ and $(t_c-t)^{(\tau-1)/\sigma-2}$
than with the theoretical curves as observed in CDMG. We show that this discrepancy occurs for the same reason as in CDMG, 
namely that $(t_c-t)$ in Fig.~\ref{Fig:PR_H_Hprime_scaling}(c) and (d)
is not small enough to reflect the $(t_c-t) \rightarrow 0$ limit used to derive the theoretical curves (see Appendix B).

For $t > t_c$, as shown in Fig.~\ref{Fig:PR_H_Hprime_scaling}(e), 
$f_2(x) \approx a_0 - a_1x^{\sigma}+a_2x^{2\sigma}$ using the estimated $a_0$, $a_1$, and $a_2$ 
is reasonable for small $x$, and $C^+$ calculated numerically using Eq.~(\ref{Eq:Cplus}) has positive values regardless of $\alpha \ll 1$.
Then, the scaling behaviors of $H(t)-H(t_c)$, $\dot{H}(t)-\dot{H}_{\infty}(t_c)$, and $\ddot{H}(t)$
fit well with the theory, as shown in Fig.~\ref{Fig:PR_H_Hprime_scaling}(f), (g), and (h), respectively.
Finally, we confirmed that the reason why the theoretical curves for $\dot{H}(t)-\dot{H}_{\infty}(t_c)$ 
and $\ddot{H}(t)$ fit well 
with the data in Fig.~\ref{Fig:PR_H_Hprime_scaling}(g) and (h), unlike the $t<t_c$ case, 
is the same as mentioned for CDMG.

In conclusion, we observed the scaling behaviors of 
$H(t)-H(t_c)$, $\dot{H}(t)-\dot{H}_{\infty}(t_c)$, and $\ddot{H}(t)$ 
predicted through the theory derived using CDMG even in the product rule.
Accordingly, we expect this approach to be applicable to various other EP models.

Finally, we mention the possibility that a discontinuous percolation transition involves a discontinuous decrease of entropy,
because entropy measures disorder such that it would decrease discontinuously as the giant cluster emerges discontinuously.
This expectation is reminiscent of the discontinuous entropy change during a discontinuous transition in thermal equilibrium systems.
In~\cite{vieira}, the maximum point of $H$ approaches the threshold 
from the left, and the maximum value of $H$ increases as $m$ increases in CDMG.
Therefore, we expect that the information entropy $H$ would decrease discontinuously 
at the threshold when a discontinuous transition occurs as $m \rightarrow \infty$ in CDMG.
We anticipate that this expectation can be clarified by extending the results of the present paper.
Furthermore, we believe that it is important to demonstrate that this is a general feature of discontinuous percolation transitions
for various definitions of entropy in percolation~\cite{vieira, entropy_latticeanimal1, entropy_latticeanimal2, hassan1, hassan2}.

\section*{Acknowledgement}
We thank Soo Min Oh for his valuable comments.
This work was supported by a National Research Foundation (NRF) of Korea grant, No. 2020R1F1A1061326.

\section*{Appendix A: Derivation of  Eq.~(\ref{Eq:Hdotscaling})}
Here, we present the detailed derivation of Eq.~(\ref{Eq:Hdotscaling}) starting from Eq.~(\ref{Eq:Hdotapprox}).
In Eq.~(\ref{Eq:Hdotapprox}), the first two terms are
\begin{widetext}
\begin{equation}
- \frac{1}{\text{ln}{2}}\left[\frac{1}{(1-t)}+\frac{1}{(1-t)^2}\int_{1}^{\infty}
n_s{\text{ln}}n_sds\right]
\approx -\frac{1}{\text{ln}{2}}\left[\frac{1}{(1-t_c)}+\frac{1}{(1-t_c)^2}\int_{1}^{\infty}
(a_0s^{-\tau}){\text{ln}}(a_0s^{-\tau})ds\right]
\tag{A1}
\label{eq:A1}
\end{equation}
up to $\mathcal{O}(1)$.
The last term in Eq.~(\ref{Eq:Hdotapprox}) is
\begin{align}
&-\frac{1}{(1-t)\text{ln}2}\bigg[\int_{1}^{\infty}\dot{n}_s{\text {ln}}n_sds
+\int_1^{\infty}\dot{n}_sds\bigg] \nonumber \\
&=\frac{1}{(1-t_c+\epsilon^{\sigma})\text{ln}2}\bigg[\frac{\tau}{\sigma}
\epsilon^{\tau-\sigma-1}{\text {ln}\epsilon}\int_{\epsilon}^{\infty}x^{1-\tau}f'_1(x)dx
+\frac{1}{\sigma}\epsilon^{\tau-\sigma-1}\int_{\epsilon}^{\infty}x^{1-\tau}f'_1(x){\text {ln}}(x^{-\tau}f_1(x))dx
+\frac{1}{\sigma}\epsilon^{\tau-\sigma-1}\int_{\epsilon}^{\infty}x^{1-\tau}f'_1(x)dx\bigg].
\label{eq:A2}
\tag{A2}
\end{align}
We divide the interval of integration $[\epsilon, \infty]$ of Eq.~(\ref{eq:A2}) into two intervals $[\epsilon, \alpha]$
and $[\alpha, \infty]$ for some $\alpha \ll 1$, and use the approximations $f_1(x) \approx a_0+a_1x^{\sigma}+a_2x^{2\sigma}$
and $f'_1(x) \approx \sigma a_1 x^{\sigma-1}+2\sigma a_2x^{2\sigma-1}$
for the first interval $[\epsilon, \alpha]$ (see Sec.~\ref{sec:negativeHprime} and Fig.~\ref{Fig:EP_H_Hprime_scaling}(a)).
Then, we expand the integral terms in Eq.~(\ref{eq:A2}) as
\begin{equation}
\int_{\epsilon}^{\infty}x^{1-\tau}f'_1(x)dx
\approx \frac{a_1\sigma\epsilon^{1+\sigma-\tau}}{(\tau-\sigma-1)}
+\left[
-\frac{a_1\sigma\alpha^{1+\sigma-\tau}}{(\tau-\sigma-1)}
-\frac{2a_2\sigma\alpha^{1+2\sigma-\tau}}{(\tau-2\sigma-1)}
-\alpha^{1-\tau}f_1(\alpha)+(\tau-1)\int_{\alpha}^{\infty}x^{-\tau}f_1(x)dx
\right]
\label{eq:A3}
\tag{A3}
\end{equation}
and
\begin{equation}
\int_{\epsilon}^{\infty}x^{1-\tau}f'_1(x)\text{ln}(x^{-\tau}f_1(x))dx
\approx -\frac{\tau a_1\sigma}{(\tau-\sigma-1)}\epsilon^{1+\sigma-\tau}\text{ln}\epsilon
+\left[
-\frac{\tau a_1\sigma}{(\tau-\sigma-1)^2}
+\frac{a_1\sigma\text{ln}a_0}{(\tau-\sigma-1)}
\right]\epsilon^{1+\sigma-\tau}
\label{eq:A4}
\tag{A4}
\end{equation}
up to $\mathcal{O}(1)$ and $\mathcal{O}(\epsilon^{1+\sigma-\tau})$, respectively.
Applying the approximations in Eq.~(\ref{eq:A1}--\ref{eq:A4}) and expanding Eq.~(\ref{Eq:Hdotapprox}) up to 
$\mathcal{O}(\epsilon^{\tau-\sigma-1}\text{log}_2\epsilon)$ for $\epsilon=(t_c-t)^{1/\sigma}$, we obtain Eq.~(\ref{Eq:Hdotscaling}).

\section*{Appendix B: Derivation of the next dominant terms of $\dot{H}(t)-\dot{H}_{\infty}(t_c)$ and $\ddot{H}(t)$
to fit the data for $t < t_c$}

In this section, we expand Eq.~(\ref{Eq:Hdotapprox}) up to $\mathcal{O}(\epsilon^{\tau-\sigma-1})$ and show that the next dominant term of $\dot{H}(t)-\dot{H}_{\infty}(t_c)$ is 
$\mathcal{O}(\epsilon^{\tau-\sigma-1})$ for $\epsilon=(t_c-t)^{1/\sigma} \ll 1$. 
We then show that the theoretical curves for $\dot{H}(t)-\dot{H}_{\infty}(t_c)$ and $\ddot{H}(t)$ including the next dominant terms fit well with the data (dashed lines and symbols) in Fig.~\ref{Fig:EP_H_Hprime_scaling}(c) and (d) for CDMG as well as in Fig.~\ref{Fig:PR_H_Hprime_scaling}(c) and (d) for the product rule.

For the first two terms in Eq.~(\ref{Eq:Hdotapprox}), we can again use the approximation Eq.~(\ref{eq:A1}) here
because the next dominant term in the expansion is $\mathcal{O}(\epsilon^{\sigma})$,
which can be ignored in this case. For the last term in Eq.~(\ref{Eq:Hdotapprox}), the first integral term in Eq.~(\ref{eq:A2}) can also be 
approximated using Eq.~(\ref{eq:A3}) here because the next dominant term in the expansion is $\mathcal{O}(\epsilon^{1+2\sigma-\tau})$,
which can be ignored. The only part that needs to be corrected is Eq.~(\ref{eq:A4}), 
which is the expansion of the second integral term in Eq.~(\ref{eq:A2}).
We expand the second integral term as
\begin{align}
\int_{\epsilon}^{\infty}x^{1-\tau}f'_1(x)\text{ln}(x^{-\tau}f_1(x))dx
&\approx -\frac{\tau a_1\sigma}{(\tau-\sigma-1)}\epsilon^{1+\sigma-\tau}\text{ln}\epsilon
+\left[
-\frac{\tau a_1\sigma}{(\tau-\sigma-1)^2}
+\frac{a_1\sigma\text{ln}a_0}{(\tau-\sigma-1)}
\right]\epsilon^{1+\sigma-\tau}
\nonumber \\
&-\frac{\sigma^3(1-t_c)(2\tau-\sigma-1)}{\tau(\tau-\sigma-1)^2}C^-
+\frac{\sigma^2(1-t_c)\textrm{ln}2}{(\tau-\sigma-1)}C^-_1
\tag{A5}
\label{eq:A5}
\end{align}
up to $\mathcal{O}(1)$, where $C^-$ is Eq.~(\ref{Eq:Cminus}) and
\begin{align}
C^-_1 &= \frac{1}{\textrm{ln}2}\bigg[\frac{\sigma(2\tau-\sigma-1)}{\tau(\tau-\sigma-1)}C^-
+\frac{(\tau-\sigma-1)}{(1-t_c)\sigma^2}
\bigg\{-\frac{a_1\sigma{\textrm{ln}}a_0}{(\tau-\sigma-1)}\alpha^{1+\sigma-\tau}
+\frac{\tau a_1\sigma}{(\tau-\sigma-1)}\alpha^{1+\sigma-\tau}{\textrm {ln}}\alpha
\nonumber \\
&+\frac{\tau a_1\sigma}{(\tau-\sigma-1)^2}\alpha^{1+\sigma-\tau}
-\frac{2\tau a_2 \sigma}{(1+2\sigma-\tau)}\alpha^{1+2\sigma-\tau}{\textrm {ln}}\alpha 
+\frac{2\tau a_2\sigma}{(1+2\sigma-\tau)^2}\alpha^{1+2\sigma-\tau}
+\frac{2a_2\sigma{\textrm {ln}}a_0}{(1+2\sigma-\tau)}\alpha^{1+2\sigma-\tau}
\nonumber \\
&+\frac{a_1^2\sigma}{a_0}\frac{\alpha^{1+2\sigma-\tau}}{(1+2\sigma-\tau)}
+\tau f_1(\alpha)\alpha^{1-\tau}{\textrm{ln}}\alpha
+\tau(1-\tau)\int_{\alpha}^{\infty}f_1(x)x^{-\tau}{\textrm {ln}}x dx
+\alpha^{1-\tau}f_1(\alpha)-\alpha^{1-\tau}f_1(\alpha){\textrm {ln}}f_1(\alpha)
\nonumber \\
&+(\tau-1)\int_{\alpha}^{\infty}x^{-\tau}f_1(x){\textrm {ln}}f_1(x)dx
+\int_{\alpha}^{\infty}f_1(x) x^{-\tau}dx
\bigg\}\bigg].
\tag{A6}
\label{eq:A6}
\end{align}
We note that the last two terms added in Eq.~(\ref{eq:A5}) are $\mathcal{O}(1)$.
We apply the approximations in Eqs.~(\ref{eq:A1}--\ref{eq:A3}) and (\ref{eq:A5})
to expand Eq.~(\ref{Eq:Hdotapprox}) up to $\mathcal{O}(\epsilon^{\tau-\sigma-1})$, and as a result obtain
\begin{align}
\dot{H} &\approx 
\frac{1}{\text{ln}2}\bigg[-\frac{1}{(1-t_c)}-\frac{a_0{\text {ln}}a_0}{(1-t_c)^2(\tau-1)}
+\frac{\tau a_0}{(1-t_c)^2(\tau-1)^2}
+\frac{a_1(1+\text{ln}a_0)}{(1-t_c)(\tau-\sigma-1)}
- \frac{\tau a_1}{(1-t_c)(\tau-\sigma-1)^2}\bigg]
\nonumber \\
&+C^-\frac{\sigma}{(\tau-\sigma-1)}(t_c-t)^{\frac{(\tau-1)}{\sigma}-1}\text{log}_2(t_c-t)
+\bigg[C^-_1-\frac{\sigma}{(\tau-\sigma-1)}
\frac{C^-}{\textrm{ln}2}\bigg]
\frac{\sigma}{(\tau-\sigma-1)}
(t_c-t)^{\frac{(\tau-1)}{\sigma}-1}.
\tag{A7}
\label{eq:A7}
\end{align}
Finally, the expansion of $\ddot{H}$ up to $\mathcal{O}(\epsilon^{\tau-2\sigma-1})$ 
can be obtained by differentiating Eq.~(\ref{eq:A7}) as
\begin{align}
\ddot{H} &\approx 
-C^-(t_c-t)^{\frac{(\tau-1)}{\sigma}-2}{\textrm{log}}_2(t_c-t)
-C^-_1(t_c-t)^{\frac{(\tau-1)}{\sigma}-2}.
\tag{A8}
\label{eq:A8}
\end{align}
Therefore, the next dominant terms of $\dot{H}(t)-\dot{H}_{\infty}(t_c)$
and $\ddot{H}(t)$ for $\epsilon = (t_c-t)^{1/\sigma} \ll 1$ 
are $\mathcal{O}(\epsilon^{\tau-\sigma-1})$ 
and $\mathcal{O}(\epsilon^{\tau-2\sigma-1})$, respectively.
\end{widetext}

\begin{figure}[t!]
\includegraphics[width=1.0\linewidth]{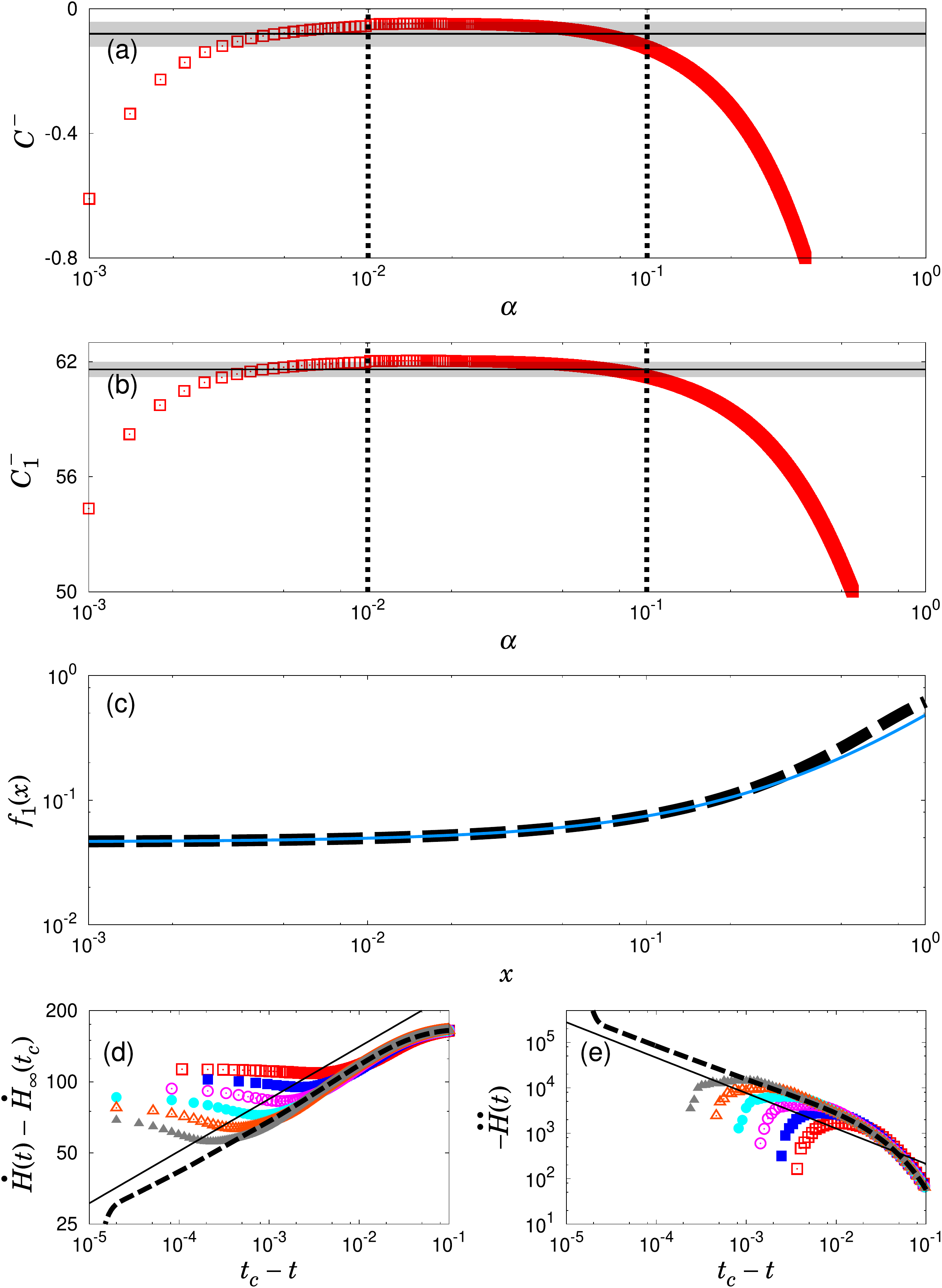}
\caption{
Fitting modified theoretical curves to the CDMG data for $t < t_c$.
(a) $C^-(\alpha)$ $(\square)$ calculated using Eq.~(\ref{Eq:Cminus})
with the numerically obtained $f_1(x)$ described in Fig.~\ref{Fig:EP_polylog_H_Hprime}(a).
The shaded area represents the maximum and minimum of $C^-(\alpha)$ 
over $10^{-2} \leq \alpha \leq 10^{-1}$ (within the dotted lines), and the straight line represents
the midrange.
We estimate that the exact value of $C^-$ is within the y range of the shaded area $-0.08 \pm 0.04$.
(b) $C^-_1(\alpha)$ $(\square)$ calculated using Eq.~(\ref{eq:A6}) with the numerically obtained $f_1(x)$ described in Fig.~\ref{Fig:EP_polylog_H_Hprime}(a). The shaded area represents the maximum and minimum of $C^-_1(\alpha)$ 
over $10^{-2} \leq \alpha \leq 10^{-1}$ (within the dotted lines), and the straight line represents
the midrange.
We estimate that the exact value of $C^-_1$ is within the y range of the shaded area $61.6 \pm 0.4$.
(c) Enlarged figure of the numerically obtained $f_1(x)$ (dashed line) and the approximation 
$f_1(x) \approx a_0 + a_1x^{\sigma} + a_2x^{2\sigma}$ (solid line) in Fig.~\ref{Fig:EP_polylog_H_Hprime}(a).
(d) The solid line $\propto (t_c-t)^{(\tau-1)/\sigma-1}{\textrm {log}}_2(t_c-t)
+[C^-_1/C^- - \sigma/(\tau-\sigma-1){\textrm {ln}}2](t_c-t)^{(\tau-1)/\sigma-1}$ 
is drawn to fit the data in Fig.~\ref{Fig:EP_H_Hprime_scaling}(c).
(e) The solid line $\propto -(t_c-t)^{(\tau-1)/\sigma-2}\textrm{log}_2(t_c-t)
-(C^-_1/C^-)(t_c-t)^{(\tau-1)/\sigma-2}$ 
is drawn to fit the data in Fig.~\ref{Fig:EP_H_Hprime_scaling}(d).
(d,e) We use $C^-=-0.08$ and $C_1^-=61.6$ estimated as the exact values in (a,b)
to draw the solid lines.
}
\label{Fig:EP_H_Hprime_scaling_correction}
\end{figure}

In Fig.~\ref{Fig:EP_H_Hprime_scaling_correction}, we check that the modified theoretical equations, Eqs.~(\ref{eq:A7}) and (\ref{eq:A8}), 
fit well with the data. In the equations, the exact values 
of $C^-$, $C_1^-$ without the approximation using $\alpha \ll 1$ 
should be derived by taking the limit as $\alpha \rightarrow 0$ after substituting the closed form
of $f_1(x)$ into Eqs.~(\ref{Eq:Cminus}) and~(\ref{eq:A6}), respectively.
However, we use $f_1(x)$ obtained numerically instead of the closed form of $f_1(x)$ in this paper. 
Therefore, in Fig.~\ref{Fig:EP_H_Hprime_scaling_correction}(a)--(c), we estimate the midrange of each of the calculated $C^-$, $C_1^-$
in the intermediate range of $\alpha$ $(10^{-2} \leq \alpha \leq 10^{-1})$,
over which the calculated value is relatively flat to approximate the exact value.
This is because the exact values of $C^-$, $C_1^-$ should be independent of $\alpha$.
Moreover, this is supported by the argument that $f_1(x)$ obtained numerically may not be exact for $x \leq \alpha$ 
due to finite size effects if $\alpha < 10^{-2}$, and also that
$f_1(x)$ may not follow the approximation $f_1(x) \approx a_0 + a_1x^{\sigma} + a_2x^{2\sigma}$ for $x \geq \alpha$
if $\alpha > 10^{-1}$.
Equation~(\ref{eq:A7}) for $\dot{H}-\dot{H}_{\infty}(t_c)$ and Eq.~(\ref{eq:A8}) for $\ddot{H}$
with the estimated values of $C^-, C_1^-$ fit well with the data as shown in Fig.~\ref{Fig:EP_H_Hprime_scaling_correction}(d), (e).
We note that $|C_1^-/C^-| \gg 1$ such that the theoretical curves look almost like straight lines in the figures, as predicted in Fig.~\ref{Fig:EP_H_Hprime_scaling}(c), (d).

\begin{figure}[t!]
\includegraphics[width=1.0\linewidth]{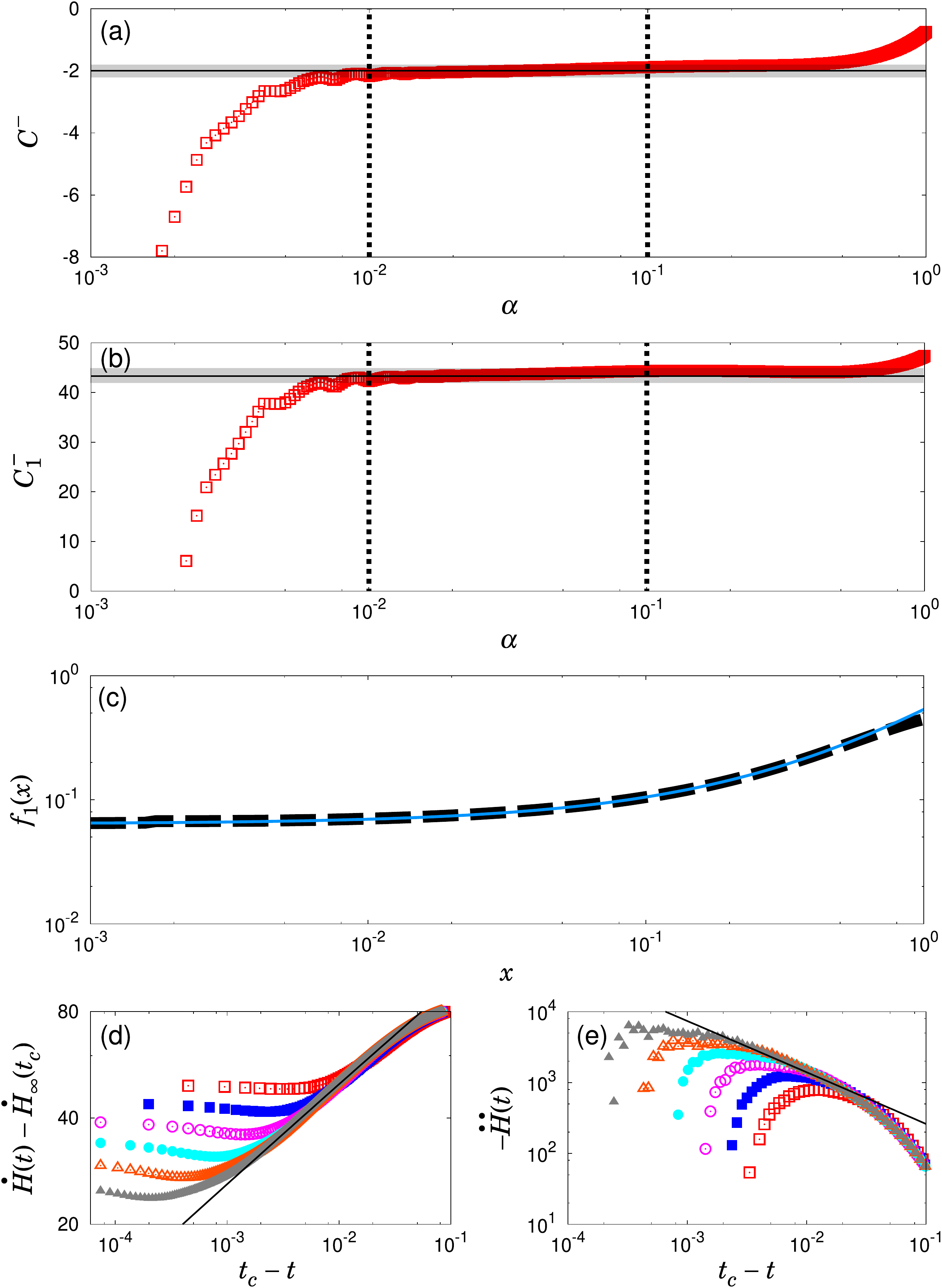}
\caption{Fitting modified theoretical curves to the product rule data for $t < t_c$.
(a) $C^-(\alpha)$ $(\square)$ calculated using Eq.~(\ref{Eq:Cminus})
with the $f_1(x)$ obtained from simulation as described in Fig.~\ref{Fig:PR_H_Hprime_scaling}(a).
The shaded area represents the maximum and minimum of $C^-(\alpha)$ 
over $10^{-2} \leq \alpha \leq 10^{-1}$ (within the dotted lines), and the straight line represents
the midrange.
We estimate that the exact value of $C^-$ is within the y range of the shaded area $-2.0 \pm 0.2$.
(b) $C^-_1(\alpha)$ $(\square)$ calculated using Eq.~(\ref{eq:A6}) with the $f_1(x)$ 
obtained from simulation as
described in Fig.~\ref{Fig:PR_H_Hprime_scaling}(a). The shaded area represents the maximum and minimum of $C^-_1(\alpha)$ 
over $10^{-2} \leq \alpha \leq 10^{-1}$ (within the dotted lines), and the straight line represents
the midrange.
We estimate that the exact value of $C^-_1$ is within the y range of the shaded area $43.3 \pm 1.5$.
(c) Plot of the $f_1(x)$ obtained from simulation (dashed line) and the approximation 
$f_1(x) \approx a_0 + a_1x^{\sigma} + a_2x^{2\sigma}$ (solid line) 
that are described in Fig.~\ref{Fig:PR_H_Hprime_scaling}(a) and extended to $10^{-3} \leq x \leq 1$.
(d) The solid line $\propto (t_c-t)^{(\tau-1)/\sigma-1}{\textrm {log}}_2(t_c-t)
+[C^-_1/C^- - \sigma/(\tau-\sigma-1){\textrm {ln}}2](t_c-t)^{(\tau-1)/\sigma-1}$ 
is drawn to fit the data in Fig.~\ref{Fig:PR_H_Hprime_scaling}(c).
(e) The solid line $\propto -(t_c-t)^{(\tau-1)/\sigma-2}\textrm{log}_2(t_c-t)
-(C^-_1/C^-)(t_c-t)^{(\tau-1)/\sigma-2}$ 
is drawn to fit the data in Fig.~\ref{Fig:PR_H_Hprime_scaling}(d).
(d,e) We use $C^-=-2.0$ and $C_1^-=43.3$ estimated as the exact values in (a,b)
to draw the solid lines.}
\label{Fig:PR_H_Hprime_scaling_correction}
\end{figure}

In Fig.~\ref{Fig:PR_H_Hprime_scaling_correction},
we repeat the same analysis for the product rule.
In Fig.~\ref{Fig:PR_H_Hprime_scaling_correction}(a)--(c),
we estimate the exact value of $C^-$ $(C_1^-)$ for the product rule 
by taking the midrange of the calculated $C^-$ $(C_1^-)$ 
in the intermediate range of $\alpha$ $(10^{-2} \leq \alpha \leq 10^{-1})$,
over which the calculated value is relatively flat.
Then, Eq.~(\ref{eq:A7}) for $\dot{H}-\dot{H}_{\infty}(t_c)$ and Eq.~(\ref{eq:A8}) for $\ddot{H}$
with the estimated values of $C^-, C_1^-$ fit well with the data as shown 
in Fig.~\ref{Fig:PR_H_Hprime_scaling_correction}(d), (e).
Similar to CDMG, $|C^-_1/C^-| \gg 1$ such that the theoretical curves look like almost straight lines in the figures, as predicted in Fig.~\ref{Fig:PR_H_Hprime_scaling}(c), (d).

\end{document}